# Alleviating polarity-conflict at the heterointerfaces of $KTaO_3$/$GdScO_3$ polar complex-oxides

J. Thompson,[1] J. Hwang,[2] J. Nichols,[1] J. G. Connell,[1] S. Stemmer,[2] and S. S. A. Seo[1, a)]

[1] *Department of Physics and Astronomy, University of Kentucky, Lexington, KY 40405, USA*
[2] *Materials Department, University of California, Santa Barbara, CA 93106, USA*
[a)] E-mail: a.seo@uky.edu

## Abstract

We have synthesized and investigated the heterointerfaces of $KTaO_3$ (KTO) and $GdScO_3$ (GSO), which are both polar complex-oxides along the pseudo-cubic [001] direction. Since their layers have the same, conflicting net charges at interfaces, i.e. $KO(-1)$/$ScO_2(-1)$ or $TaO_2(+1)$/$GdO(+1)$, forming the heterointerface of KTO/GSO should be forbidden due to strong Coulomb repulsion, the so-called *polarity conflict*. However, we have discovered that atomic reconstruction occurs at the heterointerfaces between KTO thin-films and GSO substrates, which effectively alleviates the polarity conflict without destroying the hetero-epitaxy. Our result demonstrates one of the important ways to create artificial heterostructures from polar complex-oxides.

The polarity of materials and their electrostatic boundary conditions are key factors to create unprecedented electronic and magnetic properties in complex-oxide heterostructures. For example, the discontinuous polarity at the heterointerface between polar LaAlO$_3$ (LAO) and non-polar SrTiO$_3$ (STO)[1,2] has resulted in confined electrons at the interface to form a two-dimensional electron gas (2DEG),[3,4] which exhibits intriguing properties such as metal-insulator transitions,[5] colossal capacitance,[6,7] and the coexistence of superconductivity and magnetism.[8] These phenomena are thought to originate from electron-transfer that prevents the electric potential from diverging within the polar layer, the so-called 'polar catastrophe'.[1] Here, we address a simple but important question: "What happens at heterointerfaces where two different polar complex oxides meet?" As a model system, we have investigated the heterointerfaces of KTaO$_3$ (KTO) and GdScO$_3$ (GSO), which are both polar complex-oxides along the pseudo-cubic [001] direction. Since their layers have the same, conflicting net charges at interfaces, i.e. KO(–1)/ScO$_2$(–1) or TaO$_2$(+1)/GdO(+1), forming the heterointerface of KTO/GSO should be forbidden due to the 'polarity conflict' resulting from strong Coulomb repulsion. However, we have discovered that atomic reconstruction occurs at the heterointerfaces between KTO thin-films and GSO substrates, which effectively alleviates the polarity conflict without destroying the hetero-epitaxy. Our results demonstrate an important way to create artificial heterostructures from polar complex-oxides.

There are two possible configurations of heterointerfaces between KTaO$_3$ (KTO) and GdScO$_3$ (GSO) along the pseudo-cubic [001] direction. Because the valence states of K$^+$, Ta$^{5+}$, Gd$^{3+}$, and Sc$^{3+}$ are stable, the KO (GdO) layers have a net charge of –1 (+1) and the TaO$_2$ (ScO$_2$) layers have a net charge of +1 (–1), respectively. The net charge of –1 (+1) means one electron (hole) per unit-cell square lattice in a simple ionic picture. What is controversial here is that the two adjacent atomic layers at the heterointerfaces, i.e. KO(–1)/ScO$_2$(–1) (Fig. 1 (a)) and TaO$_2$(+1)/GdO(+1) (Fig. 1 (b)), have the same net charge, in which one can expect unstable interfacial states due to strong Coulomb repulsion. Note that this so-called "*polarity conflict*", i.e. the strong electrostatic Coulomb repulsion between two polar materials at their interfaces, occurs regardless of the termination layers of KTO and GSO (Fig. 1). Hence, one may expect that the polarity conflict would result in forbidden growth of epitaxial KTO thin-films on GSO substrates and every I-V and III-III complex-oxide heterostructure. However, here we show that high-quality KTO thin-films can be grown epitaxially on atomically flat GSO substrates even with the anticipated polarity conflict at the heterointerfaces.

Figure 2 shows a few possible ways to avoid the polarity conflict at the heterointerfaces of KTO and GSO, as well as any I-V and III-III complex-oxide heterostructures. One way is to introduce a rock-salt interfacial structure of (K,Gd)O (Fig. 2 (a)), which is commonly observed in the Ruddlesden-Popper phases. Since each KO and GdO layer has a net charge of $(-1)$ and $(+1)$, respectively, the polar nature of the heterostructure can be conserved. Another way to alleviate the conflict is through the presence of defects such as oxygen vacancies (Fig. 2 (b)) or interstitial oxygen ions (Fig. 2 (c)) at the heterointerface, which provide the necessary additional charge. A more complicated resolution is to introduce an atomically mixed layer such as an interfacial bi-layer of $K_xGd_{1-x}O/Ta_ySc_{1-y}O_2$. If $x \geq 0.5$ and $x = y + 0.5$, then this interfacial bi-layer will have a net charge of $(-1)$, which will conserve the overall polarity of the system, as shown in Fig. 2 (d). For example, a bi-layer with quarter-filled Gd and Ta ions, i.e. $K_{0.75}Gd_{0.25}O/Ta_{0.25}Sc_{0.75}O_2$ ($x = 0.75$, $y = 0.25$), results in an overall net charge of $(-1)$. Complete absence of either $Gd^{3+}$ or $Ta^{5+}$ ions, i.e. $KO/Ta_{0.5}Sc_{0.5}O_2$ ($x = 1$, $y = 0.5$) or $K_{0.5}Gd_{0.5}O/ScO_2$ ($x = 0.5$, $y = 0$), will yield a net charge of $(-1)$ as well. In the following paragraphs, our experimental investigations show that the polarity conflict at the heterointerfaces between KTO and GSO is effectively resolved by forming an interfacial bi-layer of $K_xGd_{1-x}O/Ta_ySc_{1-y}O_2$ with negligible influence from interfacial defects.

We have grown epitaxial KTO thin films (30-50 nm in thickness) on atomically flat GSO $(110)_o$ single crystal substrates using pulsed laser deposition (PLD). Bulk KTO is a cubic perovskite with a lattice parameter of $a$ = 3.989 Å,[9] whose lattice mismatch with GSO (pseudo-cubic lattice, 3.967 Å) is only –0.55 % (slight in-plane compressive strain on KTO thin-films). Such a good lattice match is an ideal condition for coherent, epitaxial growth of complex-oxide thin films. While bulk KTO is an incipient ferroelectric,[10] recent studies of KTO have revealed interesting ferromagnetism at the interfaces of KTO/STO[11] and the formation of a 2DEG at KTO surfaces.[12] The PLD growth conditions were a substrate temperature of 700 °C, an oxygen partial pressure of 100 mTorr, and a laser (KrF excimer, $\lambda$ = 248 nm) fluence of 1.6 J/cm$^2$. We used a segmented target of $KNO_3$ and KTO, in which half of the target consists of a semi-circular cold-pressed $KNO_3$ pellet and the other half a KTO single crystal.[13,14] Atomically flat GSO substrates have been prepared by annealing at 1000 °C for one hour in air.

We have grown KTO thin films on GSO substrates of various miscut angles, between 0.05° and 0.18°. Figure 3 (a) and 3 (b) show topographic images of two GSO substrates with the lowest and highest miscut angles, respectively, which are obtained with an atomic force microscope. The quality of the KTO thin film has no noticeable dependence on the substrate miscut-angle (discussed in detail in the following paragraphs). Note that supplying an excess of volatile potassium ions is one of the keys for success during the PLD growth of KTO thin films.

X-ray diffraction (XRD) shows that KTO thin films are fully-strained, and epitaxially-grown on GSO substrates. XRD $\theta$-$2\theta$ scans (Fig. 3 (c)) have revealed only the (00$l$) peaks of the KTO thin films, which confirm the [001] orientation. It is remarkable that the full-width half-maxima of rocking curve scans of the thin films ($\Delta\omega \approx 0.04$ °) are comparable to that of the GSO substrates (Fig. 3 (d)), which show the high crystallinity of our KTO thin films. A typical $\Delta\omega$ is 0.04 ° for the 110 GSO peak measured with our Goebel X-ray mirror optics. X-ray reciprocal space mapping (RSM) near the GSO $(332)_o$ diffraction peak shows that the KTO thin films are fully strained to the substrates, as shown in Fig. 4 (a). The lattice parameters of the KTO thin films from this RSM are estimated as $a = 3.963$ Å and $c = 3.994$ Å. This result of synthesizing such high-quality, fully-strained KTO thin films on GSO substrates is surprising since thin-film growth should be forbidden due to the polarity conflict between the two polar materials, as discussed above. It is possible that the polarity conflict weakens when KTO thin films are grown on high miscut-angle substrates due to the increased number of step-terraces. However, as we have mentioned above, we have tested GSO substrates with various miscut angles and high-quality thin films can be grown even on substrates with a miscut angle as low as 0.05° (Fig. 3 (a)).

To probe the microscopic structure of the questionable KTO/GSO heterointerfaces, we have measured *Z*-contrast high-resolution scanning transmission electron microscopy (STEM). Our STEM samples have been prepared by 2° wedge polishing across the heterointerface and the high angle annular dark field (HAADF) cross-sectional images are acquired with a FEI Titan STEM ($C_s = 1.2$ mm, $\alpha = 9.6$ mrad, 300 kV). Figure 4 (b) shows a *Z*-contrast STEM image, which indicates that the KTO films are of high quality and fully strained; there is no indication of misfit dislocations at the interface and the thin film, which is consistent with the XRD data. It is well known that the brightness (intensity) of the *Z*-contrast STEM image depends on the atomic number (*Z*).[15] Since there is a large difference in atomic numbers between A-site ions (K ($Z = 19$) and Gd ($Z = 64$)), as well as B-site ions (Ta ($Z = 73$) and Sc ($Z = 21$)), we can easily see that the brightest dots in the film (upper) and the substrate (lower) regions of the STEM image are Ta and Gd atoms, respectively. Note the horizontal shift of the bright columns of the atoms across the interface (◄) is seen in the STEM image since Ta atoms are at B-sites while Gd atoms are at A-sites of the perovskite ($ABO_3$) structure. Hence, the rock-salt interfacial structure (Fig. 2 (a)) is ruled out: If there were a rock-salt interfacial structure, the bright columns should have appeared straight with no horizontal shift across the interface. Moreover, we can reasonably presume that a large concentration (~ $3.2 \times 10^{14}$ $cm^{-2}$) of oxygen vacancies or interstitial oxygen ions, which are suggested mechanisms of solving the polarity-conflict in Figs. 2 (b) and 2 (c), is not present in our samples. If it were, we would have observed strain relaxation from the X-ray RSM data (Fig. 4 (a)) or misfit dislocations from the STEM data (Fig. 4 (b)). Upon closer examination of the STEM data, we have observed that an atomic reconfiguration occurs at the heterointerface, which reveals important clues about how the polarity conflict is alleviated. The high-magnification STEM image in Figure 5 (a) shows that there is a bi-layer of neighboring atoms with reduced intensities near the interface, marked with filled (►) and open (▷) triangles, compared to the Ta and Gd atoms of the regions far away from the interface. The top layer (open triangle) and the bottom layer (filled triangle) can be attributed to atomically reconstructed layers of $K_xGd_{1-x}O$ and $Ta_ySc_{1-y}O_2$ layers, respectively. The good contrast in atomic numbers between K and Gd, as well as Ta and Sc allows us to readily examine the interfacial layer using STEM intensity profiles. Figure 5 (b) shows the STEM intensity line profiles along the bi-layer. While it is a formidable task to measure the exact atomic occupancy factor of the interfacial bi-layer, our best estimate of the interfacial layer using the STEM intensity profile is $K_{0.7}Gd_{0.3}O/Ta_{0.2}Sc_{0.8}O_2$, indicating that there are more K and Sc ions than Gd and Ta ions. In order to obtain these values for $x$ and $y$, we first performed an STEM intensity profile far away from the interface in both the KTO and GSO regions, along the different layers of KO, $TaO_2$, GdO, and $ScO_2$. Next, we performed an intensity profile along the mixed (K,Gd)O and $(Ta,Sc)O_2$ layers at the interface. Finally we made a comparison of the average intensities of each row and obtained the approximate estimates of $x = 0.7\pm0.1$ and $y = 0.2\pm0.1$. It is important to note that without supplying excessive K ions to the GSO substrate, by laser-ablating $KNO_3$ pellets, we are unable to fabricate these KTO thin-films. This step of supplying excessive K ions is particularly important during the initial deposition process. This growth condition may result in the deficiency of either $Gd^{3+}$ ions at A-sites or $Ta^{5+}$ ions at B-sites in the interfacial bi-layer due to the excessive supply of K ions and the $ScO_2$ termination of GSO substrates. Hence, the fully occupied interfacial bi-layer becomes $K_{0.7}Gd_{0.3}O/Ta_{0.2}Sc_{0.8}O_2$, which satisfies the conditions of $x \geq 0.5$ and $x = y + 0.5$ necessary to achieve a net charge of $(-1)$. Two extreme configurations of $KO/Ta_{0.5}Sc_{0.5}O_2$ and $K_{0.5}Gd_{0.5}O/ScO_2$ can give a net charge of $(-1)$ as well, but these configurations are not consistent with our STEM data. Thus, the polarity conflict in this heterointerface is effectively resolved by the formation of a bi-layer with a net charge of $(-1)$ resulting from atomic reconstruction at the heterointerface. Note that there is an alternating intensity along the $Ta_{0.2}Sc_{0.8}O_2$ interfacial layer while the $K_{0.7}Gd_{0.3}O$ layer does not show such a fluctuation. This suggests that there is an additional atomic ordering of Ta and Sc ions (B-site elements) at the heterointerface while the K and Gd ions are rather randomly mixed, which is schematically illustrated in Fig. 5 (c).

The atomically-reconstructed bi-layer formed between two polar layers can provide an unprecedented way to create intriguing electronic states at heterointerfaces. For instance, a dimensionally-confined, highly electron-doped interfacial layer can be formed at the heterointerfaces between two polar materials. As shown in the schematic diagram of Fig. 5 (c), the reconstructed, interfacial bi-layer should have a net charge of one extra electron per unit-cell due to the adjacent polar KTO and GSO layers. Note that an extra half-electron per unit-cell is created at the interface of LAO/STO polar/non-polar heterointerfaces to avoid the polar catastrophe of polar LAO layers.[1] Hence, in the KTO/GSO system, a simple electrostatic picture will ideally lead to a two-dimensional electronic state with a carrier density twice as large as observed in the LAO/STO system since there are two polar layers instead of just one. We have measured *dc*-transport properties of our samples as a function of temperature, and found them all to exhibit an insulating behavior. However, in order to further understand this heterostructure system, microscopic characterization such as local atomic positions and displacements are suggested as future studies. Moreover, theoretical investigations such as *ab initio* calculations of KTO/GSO heterostructures will shed light on how the interfacial bi-layer formation is preferential to other options such as rock-salt structures and interfacial defects.

In summary, we have shown that high quality KTO thin films can be grown on GSO substrates despite the polarity conflict of the heterointerfaces. The polarity conflict in this system is resolved by the formation of a reconstructed bi-layer at the heterointerface, whose net charge is (−1) per unit-cell. Our observations suggest that two-dimensionally confined states with high electron densities can be created at the heterointerfaces between two polar complex-oxides, which may result in unprecedented, intriguing physical properties.

We thank Jasminka Terzic and Justin Woods for their help preparing the $KNO_3$ pellets. This research was supported in part by the National Science Foundation through Grant No. EPS-0814194 (the Center for Advanced Materials) and by a grant from the Kentucky Science and Engineering Foundation as per Grant Agreement #KSEF-148-502-14-328 with the Kentucky Science and Technology Corporation. The work at UCSB was supported by the MRSEC Program of the National Science Foundation under Award No. DMR-1121053.

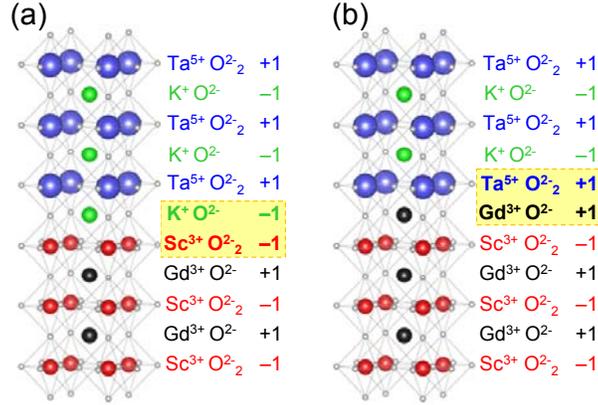

**FIG 1.** Schematic diagrams of two possible configurations of KTO/GSO heterointerface. (a) $ScO_2$ (−1) terminated GSO substrate with the first film layer of KO (−1), (b) GdO (+1) terminated GSO substrate with the first film layer of $TaO_2$ (+1).

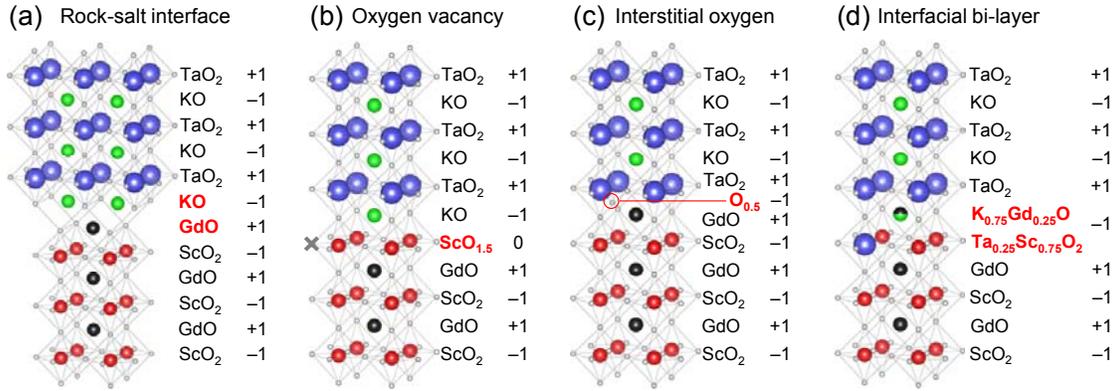

**FIG 2.** Examples of alleviating the polarity-conflict of KTO/GSO heterointerfaces. (a) The formation of a rock-salt interfacial layer. Introducing (b) 0.5 oxygen vacancies per unit-cell area of $ScO_2$ layer or (c) 0.5 interstitial oxygen ions per unit-cell area (sheet density ≈ $3.2 \times 10^{14}$ cm$^{-2}$). (d) The formation of interfacial bi-layer $K_xGd_{1-x}O/Ta_ySc_{1-y}O_2$ with $x = 0.75$ and $y = 0.25$, which gives a net charge of (−1). Any conditions satisfying $x \geq 0.5$ and $x - y = 0.5$ will yield a net charge of (−1).

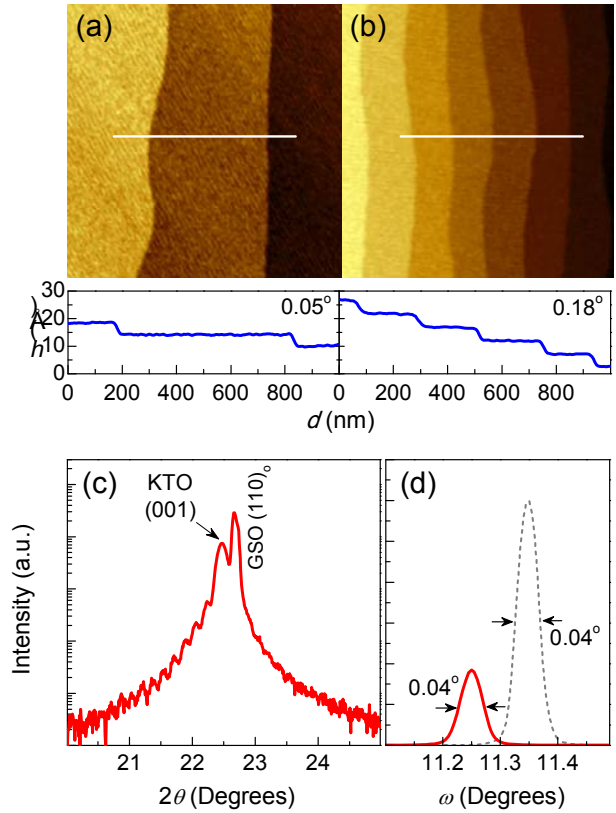

**FIG 3.** Substrate miscut angles and X-ray diffraction. Atomic force microscope topographic images of two different GSO substrates with their corresponding line profiles (white lines) of miscut angles (a) 0.05° and (b) 0.18°. (c) XRD $\theta$-$2\theta$ scan around a KTO (001) thin-film peak. (d) Rocking curves around the KTO (001) thin-film and the GSO (110)$_o$ substrate peaks.

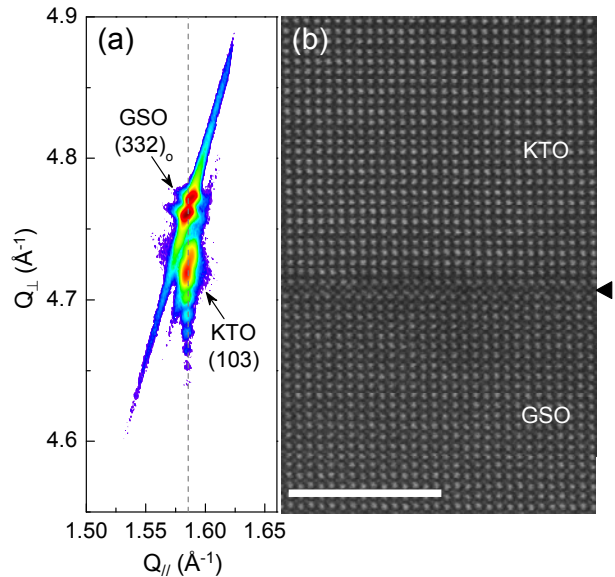

**FIG 4.** X-ray RMS and STEM data. (a) X-ray RSM around the GSO (332)$_o$ plane. The vertical dashed line indicates that the KTO film is fully strained to the GSO substrate. (b) HAADF cross-sectional STEM image of the KTO/GSO heterointerface. The white line is a 5 nm scale bar. The heterointerface between KTO and GSO is marked by a triangle (◄).

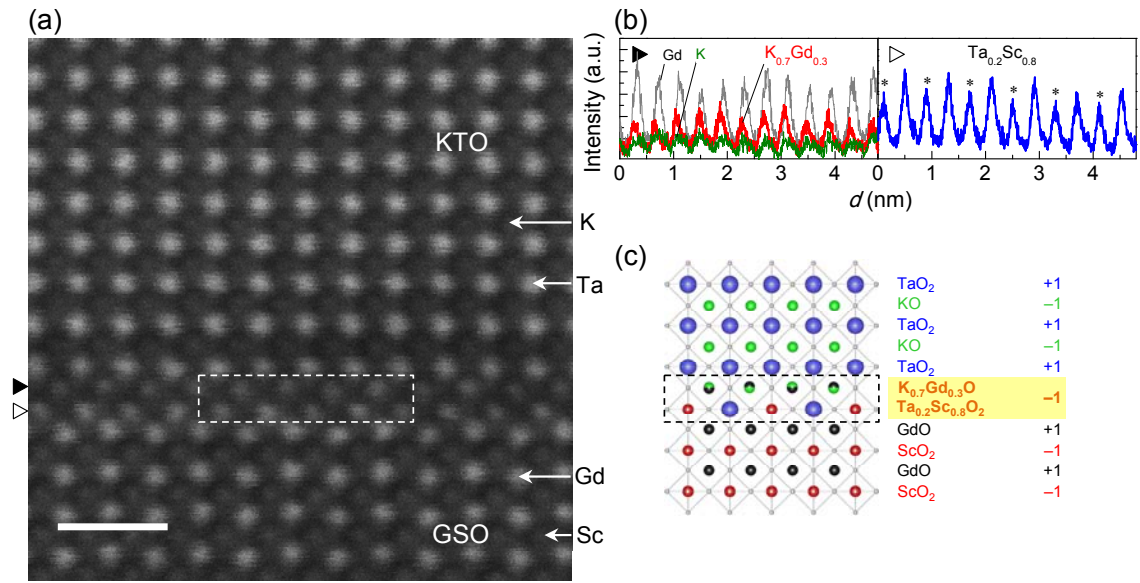

**FIG 5.** The configuration of interfacial bi-layer. (a) High-magnification STEM image of the KTO/GSO heterointerface. The white line is a 1 nm scale bar. (b) Line profiles of the bi-layers at the heterointerface. The solid and open triangles indicate the locations of the profiles in (a). The asterisks (∗) indicate reduced intensities with Ta-deficient atomic rows. (c) Schematic diagram of the reconstructed heterointerface, with the net charge of the bi-layer indicated on the right. A net charge of (–1) in the interfacial bi-layer (dashed line) maintains the overall polarity of the system.